\documentclass[sigconf]{acmart}
\AtBeginDocument{%
  }

\copyrightyear{2026}
\acmYear{2026}
\setcopyright{cc}
\setcctype{by}
\acmConference[SIGIR '26]{Proceedings of the 49th International ACM SIGIR Conference on Research and Development in Information Retrieval}{July 20--24, 2026}{Melbourne, VIC, Australia}
\acmBooktitle{Proceedings of the 49th International ACM SIGIR Conference on Research and Development in Information Retrieval (SIGIR '26), July 20--24, 2026, Melbourne, VIC, Australia}
\acmDOI{10.1145/3805712.3809912}
\acmISBN{979-8-4007-2599-9/2026/07}

\usepackage[utf8]{inputenc}
\usepackage{textgreek}
\usepackage[ruled,vlined]{algorithm2e}
\usepackage{multirow}
\usepackage{subcaption}
\usepackage{svg}
\usepackage{colortbl}
\usepackage{nicefrac}
\usepackage{algorithmic}

\usepackage{xspace}
\newcommand{\ourmethod}{\text{A2G-DiffRec}\xspace}
\newcommand{\inlinehead}[1]{\noindentparagraph{\textbf{#1}}}

\begin{document}

\title{Adaptive Autoguidance for Item-Side Fairness in Diffusion Recommender Systems}

\author{Zihan Li}
\orcid{0009-0000-9346-5058}
\affiliation{  
  \department{Institute of Computational Perception}
  \institution{Johannes Kepler University Linz}
  \city{Linz}
  \country{Austria}
}
\email{zihan.li@jku.at}

\author{Gustavo Escobedo}
\orcid{0000-0002-4360-6921}
\affiliation{%
  \department{Institute of Computational Perception}
  \institution{Johannes Kepler University Linz}
  \city{Linz}
  \country{Austria}
}
\email{gustavo.escobedo@jku.at}

\author{Marta Moscati}
\orcid{0000-0002-5541-4919}
\affiliation{%
  \department{Institute of Computational Perception}
  \institution{Johannes Kepler University Linz}
  \city{Linz}
  \country{Austria}
}
\email{marta.moscati@jku.at}

\author{Oleg Lesota}
\orcid{0000-0002-8321-6565}
\affiliation{%
  \department{Institute of Computational Perception}
  \institution{Johannes Kepler University Linz}
  \city{Linz}
  \country{Austria}
}
\email{oleg.lesota@jku.at}

\author{Markus Schedl}
\orcid{0000-0003-1706-3406}
\affiliation{
  \department{Institute of Computational Perception}
  \institution{Johannes Kepler University Linz}
  \city{Linz}
  \country{Austria}
}
\affiliation{
    \department{AI Lab}
    \institution{Linz Institute of Technology}
    \city{Linz}
    \country{Austria}
}
\email{markus.schedl@jku.at}

\renewcommand{\shortauthors}{Zihan Li, Gustavo Escobedo, Marta Moscati, Oleg Lesota, and Markus Schedl}

\begin{abstract}
Diffusion recommender systems achieve strong recommendation accuracy but often suffer from popularity bias, resulting in unequal item exposure. To address this shortcoming, we introduce \textbf{\ourmethod}, a diffusion recommender that incorporates adaptive autoguidance, where the main model is guided by a less-trained version of itself.
Instead of using a fixed guidance weight, \ourmethod learns to adaptively weigh the outputs of the main and weak models during training, supervised by a fairness-aware regularization that promotes balanced exposure across items with different popularity levels. 
Experimental results on three public datasets show that \ourmethod is effective in enhancing item-side fairness at a marginal cost of accuracy reduction compared to existing guided diffusion recommenders and other non-diffusion baselines. 
\end{abstract}

%
%


%

\maketitle

\section{Introduction and Background}
Recommender systems (RSs) are widely deployed 
to enhance user experience through personalized content delivery.
Despite their success, the predominant focus on their accuracy has raised growing concerns about fairness~\cite{jin2023fairness,deldjoo2022fairness,ekstrand2018coolkids_fairness}. A key issue that contributes
to unfairness in RSs is popularity bias, where recommendation lists are dominated by popular (head) items~\cite{Schedl2025}. This can be harmful, particularly on the item side, as long-tail items~\cite{carnovalini25popularity} receive disproportionately low exposure, reinforcing existing inequalities~\cite{abdollahpouri2017popbias,abdollahpouri2020connection_popularity}.

Existing methods to promote item-side fairness typically fall into two categories: suppressing over-exposed popular items or promoting under-represented tail items. However, these approaches often focus on only one of these directions, which may fail to address the inherent skewness in the popularity distribution~\cite{carnovalini25popularity}. 

Diffusion models have recently demonstrated strong performance in RSs~\cite{wang2023diffrec,chen-conddiff-2025}, yet their iterative denoising process may amplify skewness inherent in user--item interactions, leading to unfair outcomes~\cite{malitesta25fair}. 
To address fairness, guidance mechanisms have been introduced in diffusion-based RSs~\cite{yang25advfairdiffrec,diff-fair-rec-2024}, which guide the diffusion with statically defined weights, overlooking the fact that providing an optimal guidance weight at each step across the denoising process can improve performance~\cite{kynk24guidance,shen24guidance}. Although adaptive guidance mechanisms have been proposed to achieve fair image generation~\cite{kang2025fairgen}, they remain unexplored for fairness-aware RSs. Moreover, existing work in diffusion RSs focuses on user--side fairness, leaving item-side fairness insufficiently addressed.

To address these challenges, we present an Adaptive Autoguidance Diffusion Recommender System (\textbf{\ourmethod}), a diffusion-based RS that promotes item-side fairness while maintaining competitive accuracy. 
The main novelty of our approach is adaptive autoguidance (AG)~\cite{karras2025diffusion_bad_version_aan}, alongside a fairness-aware regularization.
Originally introduced for image generation, AG guides a diffusion model using a degraded version of itself during inference. The discrepancy between the main and weak models serves as a corrective signal, indicating how the main model deviates from its weaker counterpart. Leveraging this signal, AG seeks to identify and reduce errors made by the main model, guiding the generation process toward higher-quality outputs with a fixed guidance weight. Building on this, \ourmethod replaces the fixed weight in AG and learns step-wise guidance weights during training. In addition, we propose a popularity regularization term that jointly constrains the overexposure of highly popular items and promotes underrepresented long-tail items. At inference time, the learned weights dynamically adjust the strength of autoguidance at each denoising step, steering the generation process toward a more balanced distribution of item exposure. 
The main contributions of this work are twofold: 
(1) A novel adaptive autoguidance mechanism for item-side fairness in diffusion RSs. (2) Extensive experimental evaluation\footnote{Our implementation is available at \url{https://github.com/hcai-mms/A2G-DiffRec}.} of the proposed approach, including ablation studies. Evaluation is conducted on three real-world datasets. The results show that \ourmethod achieves consistent improvements in item-side fairness while maintaining competitive recommendation accuracy. 

\section{Methodology}

Following diffusion-based recommenders~\cite{wang2023diffrec}, each user's interaction vector $\mathbf{x}_0\in {\rm I\!R}^{|I|}$ represents the user's historical interactions over an item set of $I$ items.
In the forward process, Gaussian noise is progressively added to $\mathbf{x}_0$ over $T$ timesteps, resulting in $\mathbf{x}_t$ . 
A~denoiser is trained to reverse this corruption by minimizing the reconstruction loss $\mathcal{L}_{\mathrm{base}}$. 
During sampling (inference), the denoiser iteratively removes noise from $x_t$ to reconstruct an interaction representation for recommendation.

\textbf{AG-DiffRec:}
We first introduce AG-DiffRec, which simply employs autoguidance~\cite{karras2025diffusion_bad_version_aan} to RSs and is used as one of the baselines.
At each denoising step~$t$, we employ two denoisers:
a high-quality main model $f_1$ and a weaker guiding model $f_0$ trained on the same task
but for fewer epochs. 
Given the same noisy input $\mathbf{x}_t$ (i.e., the corrupted user interaction vector), both denoisers produce estimates of the interaction vector at step $t$: $\mathbf{z}_1 = f_1(\mathbf{x}_t, t)$ and $\mathbf{z}_0 = f_0(\mathbf{x}_t, t)$.
These outputs are combined using a fixed guidance weight $w$ across all denoising steps. Hence, the standard AG formulation is given by\begin{equation}
\mathbf{z}_{\mathrm{AG}}
= w\mathbf{z}_1 + (1 - w)\mathbf{z}_0, 
\label{eq:autoguide_w}
\end{equation}

\begin{figure}[t]
  \centering  \includegraphics[width=0.99\linewidth, ]{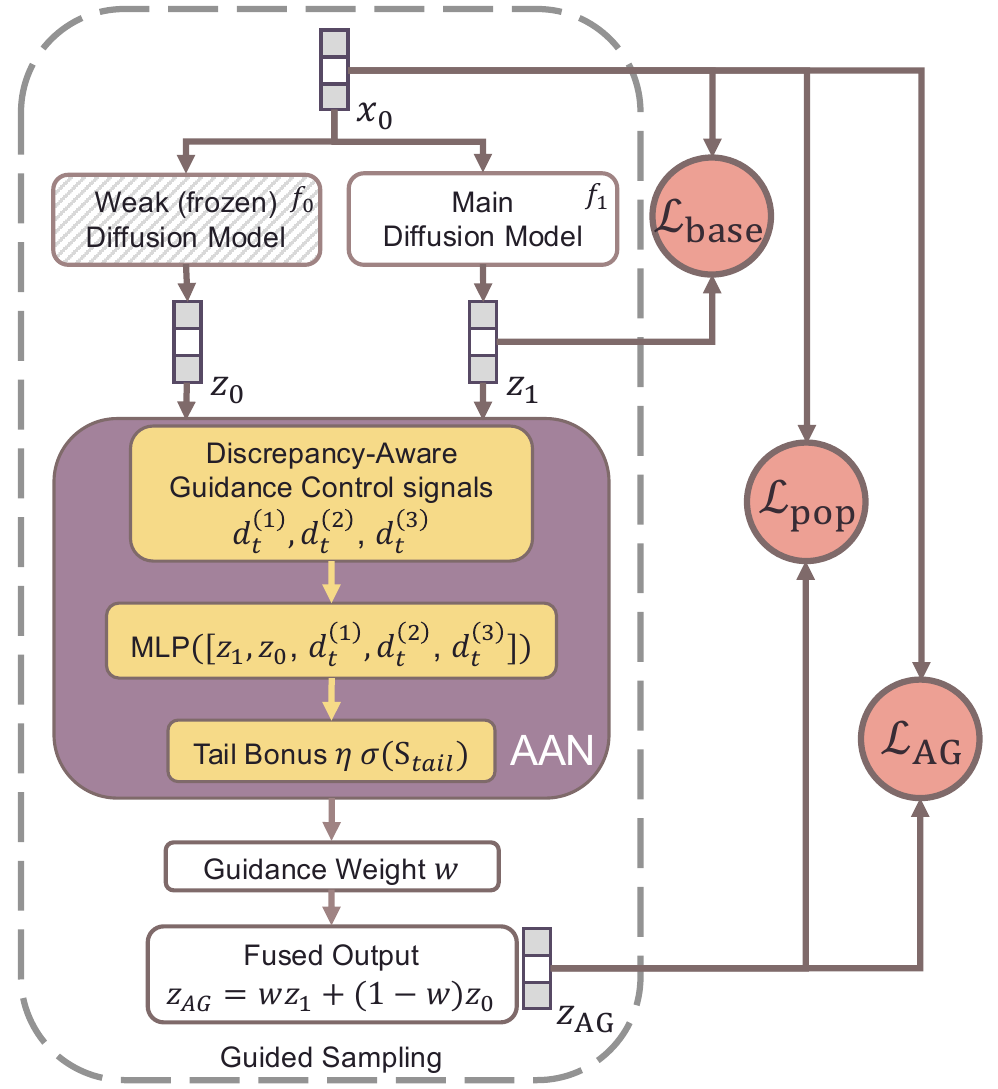}
  \caption{
Overview of \ourmethod. During training, AAN learns to produce a weight to fuse the output of the main and the weak model at each step; the same fusion is applied during sampling. 
} 
 \Description[\ourmethod framework overview]{
    A pipeline diagram of \ourmethod with training and sampling stages.
    During training, the input interaction representation is passed through a diffusion recommender with a main model and a weak model. An Adaptive Autoguidance Network (AAN) predicts a scalar weight at each denoising step, and this weight is used to fuse the outputs of the main and weak models. The fused output is used for prediction and optimization. 
    In sampling, the same main-model, weak-model, and AAN fusion mechanism is applied at each denoising step to generate the final recommendation.
  }
  \label{fig:model}
\end{figure}

\textbf{\ourmethod:}
We extend AG-DiffRec with an Adaptive Autoguidance Network (AAN), jointly optimized with the main diffusion model (see Figure~\ref{fig:model}). 
Additionally, a fairness-aware regularization is incorporated to explicitly promote balanced exposure across popularity groups. 
During sampling, the learned adaptive weights are used to fuse the main and weak model outputs at each step.

\textit{Adaptive Autoguidance Network.}
The AAN, a multilayer perceptron (MLP), learns to predict a guidance weight $w$ at each step $t$ to balance recommendation accuracy and item-side fairness.
AAN takes as input the concatenation $\mathbf{u}$ of the reconstructed vectors
$\mathbf{z}_1$ and $\mathbf{z}_0$ from the main and weak models, respectively, together with
three signals characterizing complementary aspects of the discrepancy between the two models~\cite{Lakshminarayanan17ensembles,karras2025diffusion_bad_version_aan}:
\begin{equation}
d_t^{(1)}=\|\mathbf{z}_1-\mathbf{z}_0\|_1,\hspace{0.5em}
d_t^{(2)}=\frac{\|\mathbf{z}_1\|_2}{\|\mathbf{z}_0\|_2+\varepsilon},\quad
d_t^{(3)}=H_{\tau}(\mathbf{z}_0)-H_{\tau}(\mathbf{z}_1),
\end{equation}
where $H_\tau(\mathbf{z})$ is the entropy of the temperature-scaled softmax distribution over $\mathbf{z}$, and $\tau$ is a hyperparameter to smooth the distribution.
While $\mathbf{d_t^{(1)}}$ measures the magnitude of the difference between the main model and the weak model, %
$\mathbf{d_t^{(2)}}$ captures the relative activation strength between the two models, and 
$\mathbf{d_t^{(3)}}$ quantifies the difference in prediction entropy. 
Together with the raw predictions $\mathbf{z}_1$ and 
$\mathbf{z}_0$, these components enable AAN to learn to dynamically assess the reliability and confidence of the two models’ outputs, and to predict an appropriate guidance weight that controls how to fuse the outputs.

To address item-side fairness, we apply an additional multiplicative factor when long-tail items are being recommended. Given a multi-hot vector encoding item popularity class $\mathbf{m}_{\mathrm{tail}} \in \{0,1\}^{|I|}$ where $\mathbf{m}_{\mathrm{tail},i}=1$ if item $i$ is long-tail, we compute the tail-activation score $s_{\mathrm{tail}} = \mathbf{z}_1^\top \mathbf{m}_{\mathrm{tail}}$, measuring how strongly the current recommendation $\mathbf{z}_1$ activates tail items. The final guidance weight is:
\begin{equation}
w = \left[1 + (w_{\max} - 1)\cdot\sigma\!\left(\mathrm{MLP}(\mathbf{u})\right)\right] \big(1 + \eta\, \sigma(s_{\mathrm{tail}})\big),
\end{equation}
where $w_{\max}$ is a tunable hyperparameter following the original AG design, and $\eta$ controls the tail-item bonus strength~\cite{karras2025diffusion_bad_version_aan}. 

\textit{Training Objective.} %
We use a pretrained early checkpoint of DiffRec~\cite{wang2023diffrec} as the weak model. 
The main model and the guidance network are jointly optimized (while the weak model is frozen)
via the objective: 
\begin{equation} 
\mathcal{L}
= \mathcal{L}_{\mathrm{base}}
+ \lambda_{\mathrm{AG}}\,\mathcal{L}_{\mathrm{AG}}
+ \lambda_{\mathrm{pop}}\,\mathcal{L}_{\mathrm{pop}},
\end{equation}
where $\mathcal{L}_{\mathrm{AG}} = \mathbb{E}\!\left[\|\mathbf{z}_{\mathrm{AG}} - \mathbf{x}_0\|_2^2\right]$ ensures the fused $\mathbf{z}_{\mathrm{AG}}$ accurately reconstructs $\mathbf{x}_0$,
$\mathcal{L}_{\mathrm{pop}}$ regularizes the popularity distribution 
in the top-$K$ recommendations derived from $\mathbf{z}_{\mathrm{AG}}$, and
$\lambda_{\mathrm{AG}}$ and $\lambda_{\mathrm{pop}}$ are treated as hyperparameters.
 Next, we discuss the definition of $\mathcal{L}_{\mathrm{pop}}$.
    
    
 
\textit{Popularity Regularization Term.} %
Items are 
partitioned into three bins $c \in \{{HighPop}, {MidPop}, {LowPop}\}$, where the most and least popular items, accordingly accounting for 20\% of interactions, are labeled as \textit{HighPop} and \textit{LowPop}, respectively, and the rest as \textit{MidPop}~\cite{abdollahpouri2021user_centered_pop_bias,lesota25cali}. 
For each user $u$, we define the  popularity distribution in their top-$K$ recommendations as $\mathbf{r}_u = [r_u^h, r_u^m, r_u^\ell]$, where $r_u^c = \frac{\#\text{ items in bin } c}{K}$.
During training, given a batch of users $\mathcal{B}$, we define the target distribution as 
\begin{equation}
   \mathbf{T} = [T^h, T^m, T^\ell] = \gamma\,\bar{\mathbf{H}} + (1-\gamma)\,\mathbf{Q},
\end{equation}
where $\bar{\mathbf{H}}=\frac{1}{|\mathcal{B}|}\sum_{u\in\mathcal{B}}\mathbf{H}_u$ denotes the average historical popularity distribution of batch users, and
$\mathbf{H}_u=[H_u^h,H_u^m,H_u^\ell]$ represents the proportion of user $u$’s past interactions in each popularity bin.
$\mathbf{Q}=[Q^h,Q^m,Q^\ell]$ is a reference distribution defined by hyperparameters,  
serving as a global fairness-oriented exposure prior.
The weight $\gamma = 1 - \bar{H}^h$ adaptively adjusts the balance: when historical exposure to \textit{HighPop} items is large, more emphasis is placed on $\mathbf{Q}$ to balance exposure; otherwise, regularization is relaxed.
The unified popularity regularization term is defined as:
\begin{equation}
\mathcal{L}_{\mathrm{pop}}
=
\underbrace{
[\,r^{h}_u - T^{h}\,]_{+}
}_{\text{\textit{HighPop} overexposure}}
+
\underbrace{[\,T^{\ell} - r^{\ell}_u\,]_{+}
}_{\text{\textit{LowPop} underexposure}}
+
\underbrace{[\,H(\mathbf{T}) - H(\bar{\mathbf{r}})\,]_{+}
}_{\text{Overall imbalance}},
\label{eq:unified_pop_loss}
\end{equation}
where $[\cdot]_{+}=\max(0,\cdot)$, $\bar{\mathbf{r}} = \frac{1}{|\mathcal{B}|}\sum_{u \in \mathcal{B}} \mathbf{r}_u$ is the batch-averaged recommendation distribution, and $H(\cdot) = -\sum_c p_c \log p_c$ is the Shannon entropy.
The three terms respectively penalize
(i) overexposure of \textit{HighPop} items,
(ii) underexposure of \textit{LowPop} items, and
(iii) imbalance in the overall popularity distribution.
Here, $\mathcal{L}_{\mathrm{pop}}$ influences training through importance sampling~\cite{wang2023diffrec}, not direct gradient propagation. Following DiffRec\cite{wang2023diffrec}, steps with larger loss values are sampled with higher probability. In our case, steps with more pronounced popularity bias (i.e., larger $\mathcal{L}_{\mathrm{pop}}$) are therefore sampled more frequently, increasing their contribution to the optimization of the differentiable objectives $\mathcal{L}_{\mathrm{base}}$ and $\mathcal{L}_{\mathrm{AG}}$.
Inspired by~\cite{singh18fairexp,Zehlike17faiarank}, the goal of this term is to push the model towards a better trade-off between accuracy and fairness.

\begin{algorithm}[t]
\caption{\ourmethod Sampling}
\label{alg:sampling}
\begin{algorithmic}[1]
\STATE \textbf{Input:} main model $f_1$, frozen weak model $f_0$, AAN, user interaction $\mathbf{x}_0$.
\STATE Sample noise $\boldsymbol{\epsilon} \sim \mathcal{N}(\mathbf{0}, \mathbf{I})$ and initialize $\hat{\mathbf{x}}_T$.
\FOR{$t = T, \ldots, 1$}
    \STATE $\mathbf{z}_1 \leftarrow f_1(\hat{\mathbf{x}}_t, t)$,\quad
           $\mathbf{z}_0 \leftarrow f_0(\hat{\mathbf{x}}_t, t)$
    \STATE $w \leftarrow \mathrm{AAN}(\mathbf{z}_1, \mathbf{z}_0, \mathbf{d}_t)$
    \STATE $\mathbf{z}_{\mathrm{AG}} \leftarrow w \mathbf{z}_1 + (1-w)\mathbf{z}_0$
    \STATE $\hat{\mathbf{x}}_{t-1} \leftarrow \boldsymbol{\mu}_\theta(\hat{\mathbf{x}}_t, \mathbf{z}_{\mathrm{AG}}, t)$~\cite{wang2023diffrec}
\ENDFOR
\STATE \textbf{Output:} $\hat{\mathbf{x}}_0$
\end{algorithmic}
\end{algorithm}

\textit{Guided Sampling. }During the sampling process, AAN takes as input $z_1$ and $z_0$ (outputs of main and weak denoisers at step $t$)
and outputs the adaptive guidance weight $w = \mathrm{AAN}(\mathbf{z}_1, \mathbf{z}_0)$ to fuse the two recommendations.
The fused estimate is then used to compute the reverse posterior mean and
update the sample~\cite{wang2023diffrec}. After all iterating steps, the final output provides item scores for recommendation.
The guided sampling is summarized in Algorithm~\ref{alg:sampling}.

\section{Experiments}
\label{sec:exp}
\inlinehead{Datasets:}
We evaluate our approach on 
MovieLens-1M (ML1M)
~\cite{harper15ml1m}, Foursquare Tokyo (FTKY)
~\cite{malitesta25fair,YangZCQ15ftky}, and Music4All-Onion (Onion)
~\cite{moscati22onion}. 
For ML1M, ratings are all set to 1 and treated as implicit feedback. 
For FTKY, Tokyo check-ins (April 2012–September 2013) are treated as implicit feedback. 
For Onion, we restrict the set of music listening events to those occurring in 2019 and consider only user–track interactions with a play count $\geq 2$, similar to~\cite{ganhoer_moscati2024sibrar,SiBraR25}. 
We remove duplicate user–item pairs, apply iterative 5-core filtering, sort each user’s interactions chronologically, and split the data into 7:1:2 train/validation/test sets. Table~\ref{tab:datasets} summarizes the characteristics of the datasets after pre-processing.


\begin{table}[t]
\centering
\setlength{\tabcolsep}{2pt} 
\small

\caption{
Statistics of the three datasets. 
}
\label{tab:datasets}
\begin{tabular}{lrrrr}
\toprule
\textbf{Dataset} & \textbf{\#Users} & \textbf{\#Items} & \textbf{\#Interactions} & \textbf{Sparsity (\%)} 
\\
\midrule
\textbf{ML1M} & 6,040 & 3,416 & 999,611 & 95.46 
\\
\textbf{Onion} & 14,726 & 49,342 & 2,354,875 & 99.68 
\\
\textbf{FTKY} & 11,485 & 31,931 & 618,309 & 99.83 
\\
\bottomrule
\end{tabular}
\end{table}

\inlinehead{Baselines:} We compare \textit{\ourmethod} and \textit{AG-DiffRec} with three sets of recommendation methods: naive baselines (\textit{Random} and \textit{Most Popular}), classical approaches (\textit{LightGCN}~\cite{he20lightgcn} and \textit{MultiVAE}~\cite{liang18multiVAE}), 
and diffusion-based methods (\textit{DiffRec}; 
\textit{C-DiffRec}~\cite{chen-conddiff-2025}, a conditional diffusion variant with self-guidance; 
and \textit{CFG-DiffRec}~\cite{buchanan24cfgdiff}, a variant with classifier-free guidance).

\inlinehead{Training Procedure:} 
All models are trained for up to 100 epochs, with early stopping based on validation Recall@20.

\inlinehead{Hyperparameter Settings:}
For \textit{MultiVAE}, \textit{LightGCN}, and \textit{DiffRec}, we follow the hyperparameter search spaces in~\cite{wang2023diffrec}.  
The same \textit{DiffRec} search space is applied to all diffusion-based methods to ensure a fair comparison.
For \textit{C-DiffRec} and \textit{CFG-DiffRec}, we additionally tune the conditional guidance scale in $(0.05, 1.0]$ and the CFG scale in $(1.0, 5.0]$, respectively.
For \textit{AG-DiffRec}, we use the best validated \textit{DiffRec} model as the main model and train a weak guiding model via early stopping at epoch $e_{\mathrm{weak}} \in \{1,\dots,10\}$.
\textit{\ourmethod} adopts the same main/weak model setup as \textit{AG-DiffRec} and further tunes
$w_{\max} \in [2.0, 4.0]$, the temperature $\tau \in [2.0, 3.0]$,
$\lambda_{\mathrm{pop}}$, and $\lambda_{\mathrm{AG}} \in [0.2, 1.0]$,
$Q^{h} \in [0.1, 0.4]$, $Q^{l} \in [0.4, 0.8]$, and the tail bonus coefficient $\eta \in [0.4, 0.8]$.

\inlinehead{Evaluation Metrics:}
 For the top $K$ recommended items, Evaluation is conducted at $K=\{10,20,50,100\}$, and we report $K=50$ in the paper, as it is commonly used to assess item-side fairness.
Accuracy is measured by normalized discounted cumulative gain (NDCG $\uparrow$), while item-side fairness is assessed using average percentage of long-tail items (APLT $\uparrow$) and exposure disparity between head and tail items ($\Delta$Exp $\downarrow$)~\cite{malitesta25fair}. The definition of long-tail items follows~\cite{malitesta25fair}.
We further include catalog coverage (Cov $\uparrow$)~\cite{carnovalini25popularity} and Gini index~\cite{carnovalini25popularity,Fleder09gini} (Gini $\downarrow$), to quantify the overall inequality of item exposure.
For each fairness metric $m \in \mathcal{M}$, we report the fairness--accuracy trade-off ($T_m$, $\uparrow$)~\cite{wang22itemfair} as
$T_m = \frac{\text{fairness improvement}}{\text{accuracy loss}},$
where fairness improvement is defined according to the direction of $m$ and accuracy loss is defined as the relative decrease in NDCG.

\section{Results and Analysis}
\subsection{Overall Performance Comparison}
\begin{table}[t]
\centering
\small
\setlength{\tabcolsep}{1pt}  
\caption{
Evaluation results.
Best and second-best values are in \textbf{bold} and \underline{underlined}, respectively, among diffusion-based methods only.
* and ** denote $p\!\le\!0.05$ and $p\!\le\!0.01$ vs.\ DiffRec (paired t-test). 
 $\Delta(\%)$ and $T_m$ report the relative improvement and fairness--accuracy trade-off of \ourmethod over DiffRec. Negative $T_m$ values indicate fairness improvements accompanied by an accuracy gain.}
\label{tab:main-comparison}
\begin{tabular}{p{0.15in}p{0.55in}ccccc}
\toprule
& &
\multicolumn{4}{c}{\textbf{Fairness}} &
\multicolumn{1}{c}{\textbf{Accuracy}} \\
\cmidrule(lr){3-6}\cmidrule(lr){7-7}
& Model &
$\Delta$Exp@50 $\downarrow$ &
Gini@50 $\downarrow$ &
Cov@50 $\uparrow$ &
APLT@50 $\uparrow$ &
NDCG@50 $\uparrow$ \\
\midrule

\multirow{10}{*}{\rotatebox{90}{ML1M}}
& Random 
& .0218
& .1324
& 1.000
& .8173
& .0021 \\

& Most Pop. 
& 1.000 
& .9783
& .0846
& .0000 
& .0924 \\

& LightGCN 
& .7730
& .7763
& .7769
& .2034 
& .1586 \\

& MultiVAE 
& .7816 
& .7884
& .7509
& .1894
& .1533 \\
\cmidrule(lr){2-7}

& DiffRec 
& .8894 
& .8405
& .5023
& .1087 
& \textbf{.1624} \\

& C-DiffRec 
& .8784
& .8322
& .5135
& .1190 
& .1594 \\

& \mbox{CFG-DiffRec}
& .8790 
& .8341
& .5111
& .1183 
& \underline{.1616} \\
& AG-DiffRec 
& \textbf{.8718 }$^{**}$ 
& \textbf{.8301}$^{**}$ 
& \textbf{.5340}$^{**}$ 
& \textbf{.1245} $^{**}$ 
& .1596$^{**}$  \\

& \textbf{\ourmethod}
& \underline{.8747}$^{*}$ 
& \underline{.8313}$^{*}$
& \underline{.5156}
& \underline{.1218}$^{*}$ 
& .1614 \\

\cmidrule(lr){2-7}
& $\Delta$ (\%)
& $+1.65$ 
& $+1.09$ 
& $+2.65$
& $+12.05$ 
& $-0.62$ \\
& $T_m$ 
& 2.68
& 1.78
& 4.30
& 19.57
& -- \\
\midrule

\multirow{10}{*}{\rotatebox{90}{FTKY}}
& Random 
& .0013
& .1320
& 1.000
& .8015
& .0003 \\

& Most Pop. 
& 1.000 
& .9984
& .0028
& .0000 
& .0617 \\

& LightGCN 
& .9050 
& .9293
& .4804
& .0904
& .1014 \\

& MultiVAE 
& .8532 
& .9240
& .2790
& .1308 
& .0931 \\
\cmidrule(lr){2-7}

& DiffRec
& \underline{.9624}
& \underline{.9655}
& .2278
& .0378
& \underline{.1073} \\

& C-DiffRec 
& .9643
& .9666
& .2249
& .0362
& \textbf{.1075} \\

& \mbox{CFG-DiffRec}
& .9622
& .9649
& \underline{.2314}
& \underline{.0381}
& .1069 \\
& AG-DiffRec 
& .9796$^{**}$ 
& .9782$^{**}$ 
& .1258$^{**}$ 
& .0209$^{**}$ 
& .1016$^{**}$  \\

& \textbf{\ourmethod}
& \textbf{.9586}$^{**}$
& \textbf{.9577}$^{**}$
& \textbf{.2456}$^{*}$
& \textbf{.0419}$^{*}$
& .1064$^{**}$ \\
\cmidrule(lr){2-7}
& $\Delta$ (\%)
& ${+0.39}$
& ${+0.81}$
& ${+7.81}$
& ${+10.85}$
& $-0.84$ \\
& $T_m$
& 0.47
& 0.96
& 9.32
& 12.93
& -- \\
\midrule
\multirow{10}{*}{\rotatebox{90}{Onion}}
& Random 
& .0009 
& .1454 
& 1.000
& .8007
& .0006 \\

& Most Pop. 
& 1.000 & .9990 & .0029
& .0000 
& .0101 \\

& LightGCN 
& .8058
& .8403
& .6390
& .1680
& .0823 \\

& MultiVAE 
& .7070
& .7622
& .7890
& .2430
& .0680 \\

\cmidrule(lr){2-7}

& DiffRec 
& .8343
& .8516
& .4939
& .1489
& \underline{.1043} \\

& C-DiffRec 
& \underline{.8342}
& .8510
& \underline{.4945}
& \underline{.1490}
& .1040 \\

& \mbox{CFG-DiffRec}
& .8353
& \underline{.8509}
& .4842
& .1488
& .1040 \\
& AG-DiffRec 
& .9328$^{**}$ 
& .9107$^{**}$ 
& .2789$^{**}$ 
& .0657$^{**}$ 
& .0887$^{**}$  \\

& \textbf{\ourmethod}
& \textbf{.8218}$^{**}$
& \textbf{.8400}$^{**}$
& \textbf{.5037}$^{*}$
& \textbf{.1607}$^{**}$
& \textbf{.1048}$^{*}$ \\
\cmidrule(lr){2-7}
& $\Delta$ (\%)
& $+1.50$
& $+1.36$
& $+1.98$
& $+7.92$
& $+0.48$ \\
& $T_m$
& -3.13
& -2.84
& -4.14
& -16.53
& -- \\

\bottomrule
\end{tabular}
\end{table}
Table~\ref{tab:main-comparison} reports the main results. 
Based on these results, we summarize the following main findings on guidance for item-side fairness:

\inlinehead{\ourmethod achieves consistent
accuracy--fairness balance.} Classical baselines achieve substantially better fairness metrics than all diffusion-based methods on ML1M and FTKY, often reducing the Gini index by 5--10 percentage points.\footnote{This observation is in line with prior work~\cite{malitesta25fair}. }
However, this fairness advantage often comes with notably lower accuracy. While C-DiffRec improves fairness on ML1M and Onion, it attains the highest accuracy while substantially degrading fairness on FTKY. CFG-DiffRec exhibits more stable behavior across datasets, but its fairness improvements remain limited in magnitude. On ML1M, AG-DiffRec achieves the strongest improvements across all fairness metrics, but at the cost of a substantial drop in accuracy, demonstrating that AG can promote fairness under certain conditions. On FTKY and Onion, AG-DiffRec performs poorly in both fairness and accuracy, often ranking last among diffusion-based models, showing strong dataset dependence and limited robustness. In contrast, \ourmethod consistently ranks among the top in fairness among diffusion baselines while maintaining competitive accuracy across datasets. 


\inlinehead{Fairness gains often come at the cost of accuracy, but this trade-off is not inevitable.} 
In most cases, fairness improvement is accompanied by a loss in accuracy, reflecting the difficulty of balancing these objectives. However,
\ourmethod achieves fairness gains without accuracy loss on Onion, showing that both goals can be aligned under certain conditions. 
Moreover, AG-DiffRec suffers declines in both fairness and accuracy objectives on FTKY and Onion. 
These observations highlight that effective fairness improvement depends on carefully designed guidance mechanisms.
\inlinehead{The effectiveness of \ourmethod varies across datasets and metrics.} 
Under the ML1M setting, which exhibits relatively low sparsity, \ourmethod achieves clear improvements in long-tail exposure, with only a 0.62\% decrease in accuracy and a 12.05\% relative improvement in APLT over DiffRec. The concurrent gains in coverage further indicate that the model effectively promotes fairness.
On FTKY, \ourmethod attains statistically significant improvements in APLT and Cov, demonstrating its ability to promote long-tail items. However, the absolute gains remain relatively limited on Gini and $\Delta$Exp.
On Onion, \ourmethod improves both accuracy and fairness, suggesting that adaptive autoguidance can enhance item-side fairness without necessarily harming ranking quality.

\subsection{Ablation Studies}

\begin{table}[t]
\centering
\setlength{\tabcolsep}{2pt} 
\caption{Ablation study 
on ML-1M. Trade-offs 
and $\mathrm{NDCG\, \Delta\%}$ are calculated with respect to DiffRec as baseline.}
\label{tab:ablation_tradeoff}

\scalebox{0.9}{
\begin{tabular}{lrrrrr}
\toprule
\textbf{Method} & 
$\text{NDCG@50}(\Delta\%)\uparrow$ &
$\text{Gini@50}\downarrow$ &
$\mathrm{T}_{\text{Gini}}$ &
$\text{APLT@50}\uparrow$ &
$\mathrm{T}_{\text{APLT}}$ \\
\midrule

DiffRec & 
0.1624 (--) & 
0.8405 & 
-- &
0.1087 & 
-- \\

\textbf{\ourmethod} & 
0.1614 (-0.62\%) & 
0.8313 & 
\textbf{1.78} &
0.1218 & 
\textbf{19.57} \\
\midrule

w/o $d_t^{(1)}$ & 
0.1595 (-1.79\%) & 
0.8279 & 
0.84 &
0.1266 & 
9.22 \\

w/o $d_t^{(2)}$ & 
0.1601 (-1.42\%) & 
\textbf{0.8261} & 
1.21 &
\textbf{0.1288} & 
13.06 \\

w/o $d_t^{(3)}$ & 
0.1609 (-0.92\%) &  
0.8339 & 
0.85 &
0.1175 & 
8.76 \\

w/o AG & 
\textbf{0.1628} (+0.25\%) & 
0.8398 & 
-0.34 &
0.1086 & 
0.37 \\

w/o $\mathcal{L}_{\mathrm{pop}}$ & 
0.1616 (-0.49\%) & 
0.8344 & 
1.47 &
0.1154 & 
12.51 \\

w/o Tail Bonus & 
0.1607 (-1.05\%) & 
0.8273 & 
1.50 &
0.1265 & 
15.64 \\
\bottomrule
\end{tabular}
}
\end{table}
To assess the contribution of each component, we conduct ablation studies on ML-1M by comparing \ourmethod with DiffRec and the following variants:

(1) \textit{w/o $d_t^{(1)}$}, (2) \textit{w/o $d_t^{(2)}$}, (3) \textit{w/o $d_t^{(3)}$}: remove the first, second, and third guidance signals in AAN, respectively;

(4) \textit{w/o AG}: disables autoguidance, keeping only the popularity regularization term;

(5) \textit{w/o $\mathcal{L}_{\mathrm{pop}}$}: removes the popularity regularization term.

(6) \textit{w/o Tail Bonus}: removes the long-tail bonus in AAN;

Table~\ref{tab:ablation_tradeoff} reports results on ML1M, summarized as:

\inlinehead{The three guidance signals are complementary.} Each signal contributes differently: removing $d_t^{(3)}$ causes the largest trade-off degradation, removing $d_t^{(1)}$ leads to the sharpest accuracy drop, and removing $d_t^{(2)}$ improves fairness but at a high accuracy cost. These results demonstrate that effective guidance requires combining multiple complementary perspectives.

\inlinehead{Popularity regularization term alone is insufficient.}
The \textit{w/o AG} variant 
achieves the highest accuracy but exhibits substantial degradation in fairness, especially for APLT, indicating that the combination of adaptive guidance and guided sampling plays an important role in improving item-side fairness.

\inlinehead{Popularity-aware training improves the fairness--accuracy trade-off.} 
Although fairness decreases and accuracy improves for the \textit{w/o $\mathcal{L}_{\mathrm{pop}}$} variant, the model still achieves a competitive trade-off relative to DiffRec. 
This result highlights that adaptive autoguidance alone contributes to fairness improvement. In contrast, \textit{w/o Tail Bonus} improves fairness at the expense of accuracy, indicating that Tail Bonus primarily contributes to balancing fairness and recommendation quality, even if it does not always optimize fairness metrics on its own. 
\section{Conclusions and Future Work}
We propose \ourmethod, a diffusion recommender with adaptive guidance and a unified popularity regularization term for balanced item exposure. 
We also introduce AG-DiffRec, the first diffusion recommender incorporating autoguidance, as a principled baseline. 
Experiments show that \ourmethod achieves consistent fairness gains with marginal accuracy loss compared with diffusion and non-diffusion baselines. 
While \ourmethod shows promising results, several directions remain for future work. First, our approach focuses on unconditional diffusion models with early checkpoints as the weak model; exploring richer degradation strategies and combining autoguidance with conditional diffusion RSs may yield additional insights. Second, we primarily address item-side fairness through popularity-based metrics; extending the framework to user-side fairness or multi-stakeholder fairness remains an open challenge. 
Third, the guidance mechanism increases computational cost during both training and sampling, and reducing this cost through faster sampling methods or variants of autoguidance~\cite{gu25insitu} remains an open challenge.
Finally, while we demonstrate fairness improvements in aggregate metrics, analyzing the fine-grained impact on specific item categories or user groups would provide deeper insights into the model's behavior.
\begin{acks}
This research was funded in whole or in part by the Austrian Science Fund (FWF): \href{https://doi.org/10.55776/COE12}{10.55776/COE12},  and  \href{https://doi.org/10.55776/DFH23}{10.55776/DFH23}, and \href{https://doi.org/10.55776/P36413}{10.55776/P36413}; and by the State of Upper Austria and the Federal Ministry of Education, Science, and Research through grant LIT-2024-13-SEE-111.
\end{acks}

\bibliographystyle{ACM-Reference-Format}
\balance
\bibliography{references}

\appendix

\end{document}